\documentclass[conference]{IEEEtran}
\IEEEoverridecommandlockouts
\usepackage[
  letterpaper,
  top=1in,
  left=0.75in,
  right=0.75in,
  bottom=0.8in
]{geometry}

\usepackage{cite}
\usepackage{textcomp}
\usepackage{xcolor}
\usepackage{amssymb}            
\usepackage{mathtools}          
\usepackage{mathrsfs}           
\usepackage{graphicx}           
\usepackage{subcaption}         
\usepackage[space]{grffile}     
\usepackage{url}                
\usepackage{lipsum}             
\usepackage{comment}
\usepackage{tikz}
\usepackage{algorithm}
\usepackage{algorithmic}

\usepackage{amsmath,amsfonts,amssymb}
\usepackage{hyperref}
\usepackage{cleveref}
\usepackage{authblk}
\usepackage{booktabs}
\usepackage[misc,geometry]{ifsym}

\title{\LARGE \bf
DNQ: Deep Nash Q-Network for Partially Observable \newline n-Player Games}

\author{%
Qintong Xie, Edward Koh, Xavier Cadet and Peter Chin%
}

\begin{document}
\maketitle
\thispagestyle{empty}
\pagestyle{empty}

\begin{abstract}
Many real-world competitive systems require multiple decision-makers to act simultaneously under shared constraints, limited information, and repeated interaction, as in auctions, resource allocation, and security competition. We study multi-turn simultaneous bidding as a controlled testbed for such problems and propose DNQ, a solver-in-the-loop equilibrium supervision framework for training bidding agents. DNQ alternates between trajectory collection, critic-based payoff estimation, equilibrium computation, and policy imitation. At each visited state, a shared critic predicts either pairwise payoff matrices or an exact $N$-player payoff tensor, an external solver computes equilibrium strategies, and the agents are trained by minimizing the KL divergence between their masked policies and the solver-derived equilibrium targets. We focus on a scalable pairwise formulation that greatly reduces equilibrium-solving cost and training time compared with the exact formulation, while the shared critic amortizes payoff learning across agents and states. Experiments compare the pairwise and exact variants using critic loss, policy entropy, bidding resource usage, and training cost, showing that the pairwise method scales to larger numbers of agents, whereas the exact method becomes computationally impractical as the joint game grows. These results illustrate the trade-off between strategic fidelity and scalability in repeated competitive environments.
\end{abstract}

\section{Introduction}\label{section:introduction}

Many real-world decision problems are inherently multi-agent: several decision-makers act simultaneously, compete over limited resources, and adapt to one another over time. This structure appears in domains such as auctions and market design \cite{krishna2009auction,milgrom2004putting,nisan2007algorithmic}, procurement and sourcing under strategic interaction \cite{ganuza2007competition}, cloud resource competition and multi-tenant allocation \cite{samimi2016combinatorial,ghodsi2011dominant}, and cybersecurity games \cite{alpcan2010network,tambe2011security,manshaei2013game}. In these settings, the value of one agent's action depends directly on the anticipated behavior of others. Such problems are sequential, strategically coupled, and often constrained by budgets, legal actions, or partial observations.

Such environments are challenging for standard reinforcement learning (RL). In single-agent RL, the environment is usually treated as stationary, whereas in multi-agent settings each learner changes the environment faced by the others. Consequently, self-play can be harder to stabilize and analyze, and improved returns do not always make it clear whether agents have learned broadly meaningful strategic behavior or have simply adapted to their co-trained opponents.

Game-theoretic equilibrium concepts offer a natural alternative. A Nash equilibrium characterizes a stable outcome in which no player benefits from unilateral deviation, making it an appealing target in repeated competitive domains. However, incorporating equilibrium reasoning into learning is computationally demanding, especially inside a sequential training loop and as the number of agents grows.

A classical connection between equilibrium reasoning and RL is \emph{Nash Q-learning} \cite{hu2003nash}, which treats each state of a Markov game as a stage game induced by learned Q-values and selects actions from a Nash equilibrium. Although conceptually attractive, this approach is difficult to scale in practice due to repeated equilibrium computation and approximate value estimation.

To study these challenges in a controlled setting, we introduce a multi-turn simultaneous bidding environment. At each turn, agents bid simultaneously under budget and legality constraints, and the outcome determines the next state. Since each visited state admits an explicit normal-form representation, equilibrium strategies can be computed directly under the valid action mask, making the environment well suited for solver-in-the-loop supervision.

On top of this environment, we propose \textbf{DNQ}, a solver-driven training pipeline that combines neural function approximation with game-theoretic supervision. DNQ uses a critic to estimate state-dependent payoffs, an external solver to compute equilibrium targets, and policy updates that minimize KL divergence to the corresponding masked equilibrium strategies. This separation between payoff estimation and equilibrium computation yields a more interpretable training process and supports controlled comparison of strategic approximations.

A central focus of the paper is the comparison between two critic formulations within DNQ. The exact critic predicts the full $N$-player payoff tensor and preserves the complete strategic interaction, but its cost scales poorly with the number of agents and actions. To improve scalability, we introduce a pairwise critic that models the game through pairwise payoff matrices. This approximation greatly reduces equilibrium-solving cost and training time while retaining useful competitive structure. A shared critic architecture further amortizes payoff learning across agents and states.

Our main contributions are as follows:
\begin{itemize}
    \item \textbf{A multi-turn bidding environment for equilibrium-guided learning.} We introduce a controlled multi-turn bidding environment with budget and legality constraints and explicit masked normal-form structure.
    \item \textbf{Solver-in-the-loop equilibrium supervision via DNQ.} We propose a training framework that uses equilibrium solvers inside the learning loop to supervise policy learning under legal action constraints.
    \item \textbf{A scalable pairwise critic.} We develop a pairwise payoff formulation and compare it with an exact critic in a shared framework, exposing the trade-off between strategic fidelity and scalability.
\end{itemize}

Overall, DNQ and the proposed environment provide a controlled and interpretable testbed for equilibrium-informed learning in repeated competitive systems.
\section{Related Work}

\textbf{Equilibrium methods in stochastic games:}
Nash equilibrium is a standard notion of strategic stability in Markov and stochastic games \cite{littman1994markov}. It provides a principled target in competitive settings, but exact equilibrium computation scales poorly with the number of agents and actions \cite{daskalakis2009complexity,do2017game}. Our work addresses this practical bottleneck by keeping equilibrium supervision in the loop while introducing a pairwise critic that reduces the cost of repeated game solving relative to exact $N$-player models.

\textbf{Multi-agent RL and opponent-dependent learning:}
Multi-agent reinforcement learning (MARL) is widely used for sequential multi-agent decision-making \cite{littman1994markov,busoniu2008comprehensive}, including cybersecurity applications \cite{liu2019machine,hammad2024deep,hu2020automated}. A core challenge is that learning agents change one another's environment during training, making policy updates difficult to interpret. Related work such as Q-Mixing \cite{smith2023learning} exploits structure across mixtures of opponent strategies by learning against pure opponents and transferring to mixed opponents through value averaging. Our approach is complementary: rather than transferring across opponent mixtures, we use solver-derived equilibrium targets to guide policy learning directly at each visited state. 

\textbf{Nash Q-learning and equilibrium-based deep RL:}
Nash Q-learning \cite{hu2003nash} is a classical attempt to combine RL with equilibrium reasoning by solving a stage-game equilibrium from learned Q-values at each state. More recent deep variants improve representation power in general-sum and partially observable games \cite{casgrain2022deep,xie2025nash}. However, these methods still face a central scalability issue: exact equilibrium-based supervision becomes increasingly expensive as the game size grows. We address this directly by comparing an exact critic with a scalable pairwise approximation that preserves solver-in-the-loop supervision while reducing training cost.

\textbf{Structured approximations and shared critics:}
A common strategy for reducing complexity in multi-agent learning is to impose structure on the interaction model. In our setting, this leads to the pairwise critic, which replaces the full joint payoff tensor with pairwise payoff matrices. This approximation targets the main computational bottleneck of solver-in-the-loop learning and is one of our central contributions. We further use a shared critic to amortize payoff estimation across agents and states, enabling direct comparison between exact and pairwise formulations within the same framework.

\textbf{Auctions, bidding, and equilibrium-guided environments:}
Auctions and bidding are natural domains for strategic reasoning \cite{krishna2009auction,milgrom2004putting,nisan2007algorithmic}, but they are often studied for economic analysis rather than as controlled benchmarks for equilibrium-guided learning. Our work contributes a multi-turn simultaneous bidding environment designed for this purpose. Because each visited state defines an explicit masked normal-form game, the environment supports direct computation of exact or approximate equilibrium targets and enables controlled study of solver-in-the-loop supervision.

\textbf{Hybrid equilibrium-informed learning:}
A broader line of work combines learning with equilibrium constraints, game solvers, or game-theoretic regularization \cite{hazra2022applications}. These methods are promising, but evaluation is often difficult when equilibrium structure is only implicit. Our work differs in coupling equilibrium-informed learning with a controlled environment where masked equilibrium targets are explicitly computable, allowing us to isolate the role of the environment, the supervision mechanism, and the exact--pairwise trade-off.
\section{Methodology}\label{section:method}

We model the interaction as a partially observable Markov game
\begin{equation}
\mathcal{G}=\langle \mathcal{S},\{\mathcal{A}_i\}_{i=1}^N,P,\{r_i\}_{i=1}^N,\gamma,\{\mathcal{O}_i\}_{i=1}^N,\{O_i\}_{i=1}^N\rangle .
\label{eq:pomg}
\end{equation}
Each agent observes a local vector $o_{i,t}\in\mathbb{R}^{2N+1}$ and selects a bid
$a_{i,t}\in\mathcal{A}_i=\{0,1,\dots,v_{\max}\}$. Legal actions depend on the current budget/value, which we encode using a binary mask $m_{i,t}\in\{0,1\}^{|\mathcal{A}_i|}$. We refer to the proposed solver-in-the-loop training framework as \textbf{DNQ} (Deep Nash Q-Network).

\subsection{Problem Setting}

We study a multi-turn simultaneous-action $N$-agent game. At each stage $t \in \{1,\dots,H\}$, each agent $i \in \{1,\dots,N\}$ selects a bid action
\begin{equation}
a_i^t \in \mathcal{A}_i(s^t) = \{0,1,\dots,v_i^t\},
\label{eq:legal_action_set}
\end{equation}
where $v_i^t$ denotes the agent's remaining budget or value at stage $t$. The global state is represented as
\begin{equation}
s^t = \big(v_1^t,\dots,v_N^t,\; b_1^{t-1},\dots,b_N^{t-1},\; u_1^{t-1},\dots,u_N^{t-1}\big),
\label{eq:global_state}
\end{equation}
where $b_i^{t-1}$ denotes the previous bid-related signal for agent $i$ and $u_i^{t-1}$ denotes a previous success indicator. The environment transition is governed by a bidding simulator, which returns a winner $w^t \in \{1,\dots,N\}$ and updates the remaining values and auxiliary state variables.

The immediate reward is defined as
\begin{equation}
r_i^t =
\begin{cases}
1, & \text{if } i = w^t,\\
0, & \text{otherwise}.
\end{cases}
\label{eq:reward}
\end{equation}
Thus, learning is driven by repeated stage-game interactions over a finite horizon $H$.

\subsection{Agent Observations and Policies}

Each agent receives a local observation composed of its own remaining value together with the global bid and success signals:
\begin{equation}
o_i^t = \phi_i(s^t) = \big(v_i^t,\; b_1^{t-1},\dots,b_N^{t-1},\; u_1^{t-1},\dots,u_N^{t-1}\big).
\label{eq:local_obs}
\end{equation}
Hence, the observation dimension for each agent is
\begin{equation}
d_{\text{obs}}^{\pi} = 1 + 2N.
\label{eq:obs_dim}
\end{equation}

Each agent is parameterized by a policy network $\pi_{\theta_i}(a_i \mid o_i)$ implemented as a multilayer perceptron with ReLU hidden activations and a softmax output layer. Since not all actions are legal at every state, we define a legality mask
\begin{equation}
m_i^t(a) =
\begin{cases}
1, & a \le v_i^t,\\
0, & \text{otherwise}.
\end{cases}
\label{eq:mask}
\end{equation}
The masked policy used for action selection is
\begin{equation}
\tilde{\pi}_{\theta_i}(a \mid o_i^t, m_i^t)
=
\frac{\pi_{\theta_i}(a \mid o_i^t)\, m_i^t(a)}
{\sum_{a'} \pi_{\theta_i}(a' \mid o_i^t)\, m_i^t(a')}.
\label{eq:masked_policy}
\end{equation}
Agents act either by sampling from $\tilde{\pi}_{\theta_i}$ during data collection or by taking the greedy action during evaluation.

\subsection{Overview of DNQ}

DNQ alternates between:
\begin{enumerate}
    \item collecting trajectories using the current agent policies,
    \item fitting a critic that predicts stage-game payoffs from the current state,
    \item solving an equilibrium of the critic-induced game,
    \item updating each policy to imitate the equilibrium strategy.
\end{enumerate}
This yields a game-theoretic policy iteration scheme in which the critic models the strategic interaction at each state and the policies are trained to match equilibrium behavior implied by that model.

\subsection{Trajectory Collection}

At the beginning of each epoch, we sample a batch of episodes from the bidding environment using the current policies. For each state $s^t$, agent $i$ forms $o_i^t = \phi_i(s^t)$, computes the legal mask $m_i^t$, and selects an action
\begin{equation}
a_i^t \sim \tilde{\pi}_{\theta_i}(\cdot \mid o_i^t, m_i^t).
\label{eq:action_sampling}
\end{equation}
The environment then returns the winner and the updated state. The collected batch contains tuples of the form
\begin{equation}
(s^t, a_1^t,\dots,a_N^t, r_1^t,\dots,r_N^t).
\label{eq:transition_tuple}
\end{equation}

\subsection{Critic Architectures}

We consider two critic parameterizations.

\subsubsection{Pairwise Critic}

In the pairwise variant, the critic predicts a zero-sum payoff matrix for each unordered pair of agents $(i,j)$:
\begin{equation}
Q_{ij}(s) \in \mathbb{R}^{A \times A},
\label{eq:pairwise_q}
\end{equation}
where $A = v_{\max} + 1$ is the maximum number of possible bid actions. Each entry $Q_{ij}(s)[a_i,a_j]$ estimates the pairwise payoff to agent $i$ when agents $i$ and $j$ choose actions $a_i$ and $a_j$, respectively.

The pairwise critic uses a shared multilayer perceptron to encode the global state $s$, followed by a separate linear output head for each pair $(i,j)$. To train the pairwise critic, we derive a pairwise target from the observed winner:
\begin{equation}
y_{ij}^t =
\begin{cases}
1, & \text{if agent } i \text{ wins at } t,\\
-1, & \text{if agent } j \text{ wins at } t,\\
0, & \text{otherwise}.
\end{cases}
\label{eq:pairwise_target}
\end{equation}
Given the sampled action pair $(a_i^t,a_j^t)$, the loss is
\begin{equation}
\mathcal{L}_{\text{pair}} =
\frac{1}{|\mathcal{D}_{\text{pair}}|}
\sum_{(s^t,\mathbf{a}^t)} \sum_{i<j}
\left(
Q_{ij}(s^t)[a_i^t,a_j^t] - y_{ij}^t
\right)^2.
\label{eq:pairwise_loss}
\end{equation}

\subsubsection{Exact Tensor Critic}

In the exact variant, the critic predicts the full normal-form payoff tensor for all agents:
\begin{equation}
Q(s) \in \mathbb{R}^{N \times A \times \cdots \times A},
\label{eq:exact_q}
\end{equation}
with one action dimension per agent. For a joint action profile $\mathbf{a}=(a_1,\dots,a_N)$, the predicted payoff to agent $i$ is
\begin{equation}
Q_i(s,\mathbf{a}).
\label{eq:exact_q_entry}
\end{equation}
This critic is trained on the observed immediate reward vector $\mathbf{r}^t = (r_1^t,\dots,r_N^t)$ using
\begin{equation}
\mathcal{L}_{\text{exact}} =
\frac{1}{|\mathcal{D}_{\text{exact}}|}
\sum_{(s^t,\mathbf{a}^t,\mathbf{r}^t)}
\frac{1}{N}
\sum_{i=1}^{N}
\left(
Q_i(s^t,\mathbf{a}^t) - r_i^t
\right)^2.
\label{eq:exact_loss}
\end{equation}

\subsection{Equilibrium Target Computation}

After obtaining critic predictions for a state $s$, we derive target policies by solving the induced game.

\subsubsection{Pairwise Equilibrium Targets}

For each pair $(i,j)$, we solve a two-player zero-sum game defined by the payoff matrix $Q_{ij}(s)$ restricted to legal actions. Let
\begin{equation}
\sigma_{ij}^{(i)} \in \Delta(\mathcal{A}_i(s)),
\qquad
\sigma_{ij}^{(j)} \in \Delta(\mathcal{A}_j(s))
\label{eq:pairwise_eq}
\end{equation}
denote the Nash equilibrium mixed strategies for agents $i$ and $j$, respectively. Each agent may participate in multiple pairwise games, so we aggregate its targets by averaging:
\begin{equation}
\bar{\sigma}_i(a)
=
\frac{1}{|\mathcal{N}(i)|}
\sum_{j \in \mathcal{N}(i)}
\sigma_{ij}^{(i)}(a),
\label{eq:pairwise_avg}
\end{equation}
where $\mathcal{N}(i)=\{j \neq i\}$ is the set of other agents. The resulting vector is normalized to obtain a valid probability distribution.

\subsubsection{Exact Equilibrium Targets}

For the exact critic, we construct an $N$-player normal-form game from the predicted tensor $Q(s)$. Illegal action profiles are discouraged by assigning a large negative payoff to agents taking illegal actions. We then compute pure Nash equilibria of the induced game using enumeration. If multiple pure equilibria are found, we average their action indicators to form a target distribution for each agent. Formally, if $\mathcal{E}(s)$ is the set of pure equilibria and $e_i(a)$ denotes the indicator that action $a$ is played by agent $i$ in equilibrium $e$, then the target is
\begin{equation}
\bar{\sigma}_i(a)
=
\frac{1}{|\mathcal{E}(s)|}
\sum_{e \in \mathcal{E}(s)} e_i(a).
\label{eq:exact_target}
\end{equation}
If no pure equilibrium exists, no exact policy target is generated for that state.

\subsection{Policy Update}

Each policy is trained to match the equilibrium-derived targets using KL divergence. Given a batch of observations, legal masks, and target distributions $\bar{\sigma}_i$, we minimize
\begin{equation}
\mathcal{L}_{\pi_i}
=
\mathbb{E}_{o_i \sim \mathcal{D}}
\left[
\sum_{a}
\bar{\sigma}_i(a)
\left(
\log \bar{\sigma}_i(a)
-
\log \tilde{\pi}_{\theta_i}(a \mid o_i, m_i)
\right)
\right].
\label{eq:policy_loss}
\end{equation}
This objective corresponds to minimizing
\begin{equation}
D_{\mathrm{KL}}(\bar{\sigma}_i \,\|\, \tilde{\pi}_{\theta_i}),
\label{eq:kl}
\end{equation}
so that the learned policy imitates the equilibrium strategy implied by the critic.

\subsection{Training Algorithm}

\begin{figure}
    \centering
    \includegraphics[width=1\linewidth]{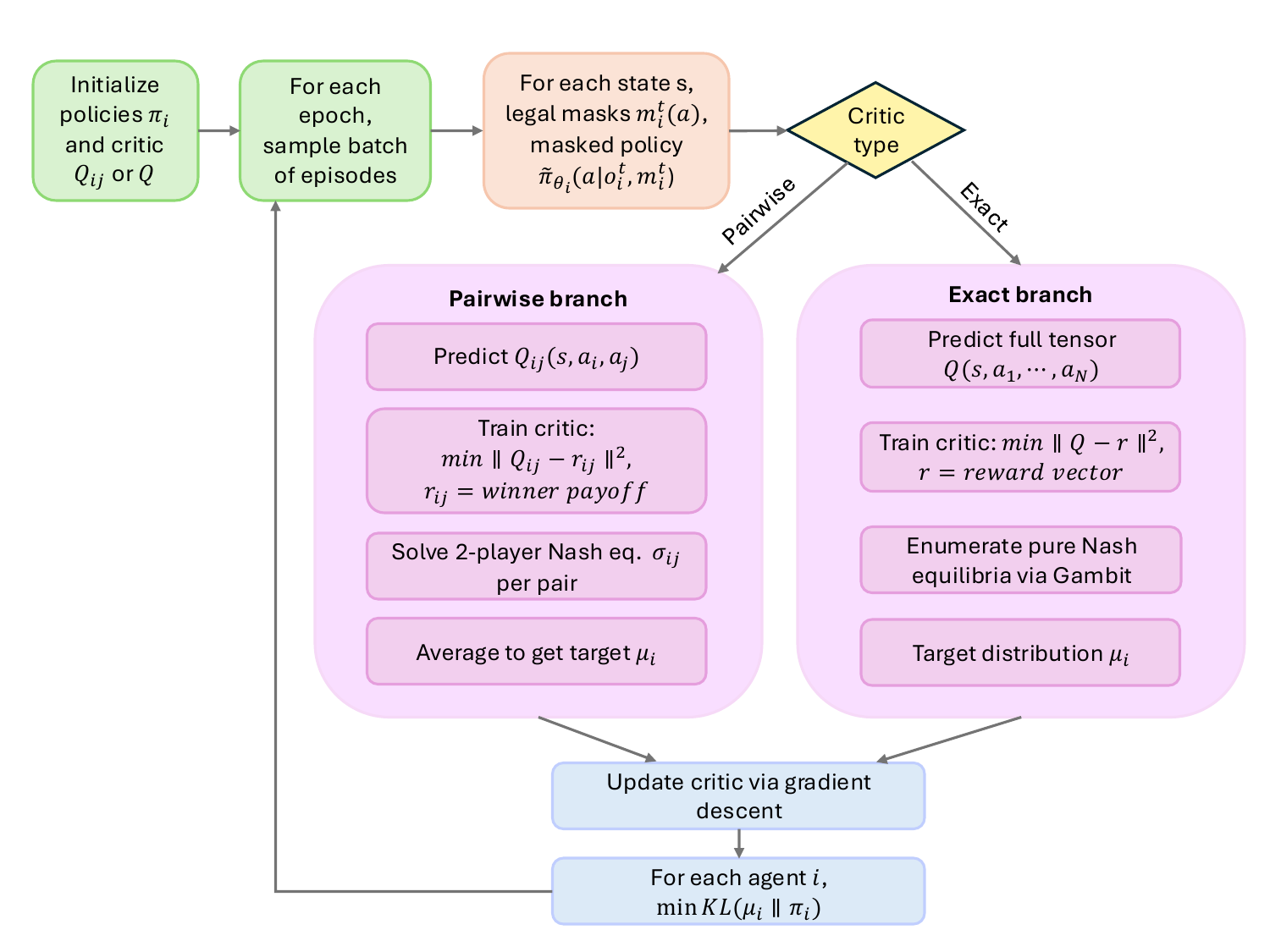}
    \caption{DNQ training pipeline. Policies are updated by minimizing KL divergence to equilibrium targets derived from either a pairwise critic or an exact tensor critic.}
    \label{fig:workflow}
\end{figure}

The overall DNQ training pipeline is shown in Figure~\ref{fig:workflow} and summarized in Algorithm~\ref{alg:dnq}. At each epoch, DNQ samples trajectories using the current policies, fits the critic to observed rewards, solves the induced pairwise or exact games, and updates the policies by minimizing KL divergence to the resulting equilibrium targets. The method therefore alternates between \emph{payoff estimation} and \emph{equilibrium imitation}.

\begin{algorithm}
\caption{DNQ: Equilibrium-Guided Multi-Agent Bidding}
\label{alg:dnq}
\begin{algorithmic}[1]
\STATE Initialize agent policies $\{\pi_{\theta_i}\}_{i=1}^N$
\STATE Initialize critic $\psi$ (pairwise or exact)
\FOR{epoch $=1,\dots,E$}
    \STATE Sample a batch of episodes using current policies
    \STATE Initialize empty target buffers for all agents
    \FOR{each visited tuple $(s,\mathbf{a},\mathbf{r})$}
        \STATE Compute legal masks $\{m_i\}_{i=1}^N$
        \IF{pairwise critic}
            \STATE Predict $\{Q_{ij}(s)\}_{i<j}$
            \STATE Accumulate pairwise critic loss using Eq.~\eqref{eq:pairwise_loss}
            \STATE Solve legal pairwise games for equilibrium strategies
            \STATE Aggregate pairwise equilibrium targets using Eq.~\eqref{eq:pairwise_avg}
        \ELSE
            \STATE Predict full payoff tensor $Q(s)$
            \STATE Accumulate exact critic loss using Eq.~\eqref{eq:exact_loss}
            \STATE Construct the induced $N$-player game
            \STATE Enumerate pure Nash equilibria and form targets using Eq.~\eqref{eq:exact_target}
        \ENDIF
        \STATE Store $(o_i,m_i,\bar{\sigma}_i)$ for each agent $i$
    \ENDFOR
    \STATE Update critic parameters by gradient descent
    \FOR{each agent $i$}
        \STATE Update $\pi_{\theta_i}$ by minimizing Eq.~\eqref{eq:policy_loss}
    \ENDFOR
\ENDFOR
\end{algorithmic}
\end{algorithm}
\section{Experiments}

\subsection{Experimental Setup}

We evaluate DNQ in a repeated bidding environment modeled as an $H$-turn simultaneous-move game with $N$ agents and maximum initial value $v_{\max}$. At the start of each episode, each agent is assigned an integer remaining budget drawn uniformly from $\{1,\dots,v_{\max}\}$. At every turn, all agents submit bids simultaneously. A bid is legal if it does not exceed the agent's current remaining budget. Among all legal bids, the highest bid wins, with ties broken uniformly at random. The winner's remaining budget is reduced by the submitted bid, while the remaining agents keep their current budgets. The environment returns the winning agent, the updated remaining budgets, the submitted bids, and a bid-status vector indicating whether each bid was illegal, legal but losing, or legal and winning.

Our experiments study two uses of DNQ. In the 2-agent setting, DNQ is instantiated only in its exact form, since the stage game reduces to a two-player matrix game whose Nash equilibrium can be computed directly. In the multi-agent settings, we compare two critic variants:
\begin{itemize}
    \item \textbf{Exact}: a full $N$-player critic that predicts the stage-game payoff tensor and computes policy targets from exact pure-strategy equilibria when available.
    \item \textbf{Pairwise}: a reduced critic that decomposes the interaction into pairwise games and forms each agent's target policy by averaging equilibria from all pairwise subgames.
\end{itemize}
In all cases, the critic is shared across agents through a common state encoder, which amortizes payoff estimation across agents and states.

The experiments are conducted in the three settings shown in Table~\ref{tab:exp_settings}: a \emph{two-agent baseline}, a \emph{three-agent comparison}, and a \emph{four-agent scaling} setting. The 2-agent and 3-agent experiments are repeated over 3 runs and we report averaged curves. These results are shown over the full training horizon, while for the 4-agent setting we report the first 80 epochs of a representative run to keep the comparison readable and to highlight the emerging runtime gap.

\begin{table}[t]
\centering
\caption{Experimental settings used to evaluate DNQ.}
\label{tab:exp_settings}
\begin{tabular}{l c c c}
\toprule
Setting & $N$ & $H$ & $v_{\max}$ \\
\midrule
Two-agent baseline & 2 & 5 & 10 \\
Three-agent comparison & 3 & 5 & 10 \\
Four-agent scaling & 4 & 5 & 15 \\
\bottomrule
\end{tabular}
\end{table}

\subsection{Evaluation Criteria}

We evaluate DNQ using four metrics:
\begin{itemize}
    \item \textbf{Critic loss}, which measures how accurately the critic fits observed stage-game rewards.
    \item \textbf{Cumulative training time}, measured directly during training and used to quantify computational efficiency.
    \item \textbf{Overall budget usage ratio}, computed as the mean fraction of initial valuation budget consumed across sampled episodes.
    \item \textbf{Mean policy entropy}, computed from the masked action distribution and averaged across agents and sampled states; lower entropy indicates more concentrated bidding behavior.
\end{itemize}

These metrics allow us to separate critic fidelity, behavioral quality, and computational cost. This is important because the exact variant models the full joint game, whereas the pairwise variant relies on a lower-dimensional approximation.

\section{Results}

\subsection{2-Agent Exact DNQ}

\begin{figure}[t]
    \centering
    \includegraphics[width=\linewidth]{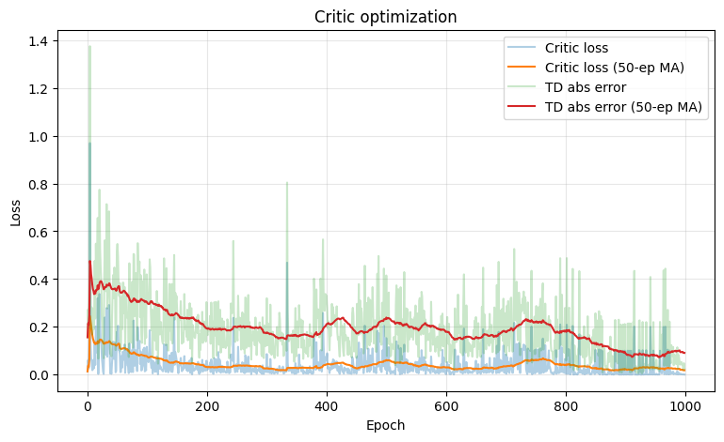}
    \caption{Training dynamics in the 2-agent exact DNQ setting $(N,H,v_{\max})=(2,5,10)$. The shared exact critic is optimized stably, and the policy becomes more structured over training.}
    \label{fig:two_agent_main}
\end{figure}

We first consider the 2-agent setting $(N,H,v_{\max})=(2,5,10)$ in Figure~\ref{fig:two_agent_main}. We use this setting as a sanity check for the solver-in-the-loop training pipeline with a shared critic.

The results show stable training dynamics. Critic loss decreases over training, and the policy becomes progressively more structured, as reflected in the entropy trend. This indicates that exact equilibrium supervision provides a usable learning signal in the simplest regime. Because the critic is shared, DNQ learns a common payoff representation rather than separate value models for each agent.

Overall, the 2-agent experiment confirms that exact DNQ behaves as intended when exact equilibrium computation is tractable. This provides a clean baseline before moving to the higher-agent settings, where scalability becomes the main issue.

\subsection{3-Agent Comparison}

\begin{figure}[t]
    \centering
    \includegraphics[width=\linewidth]{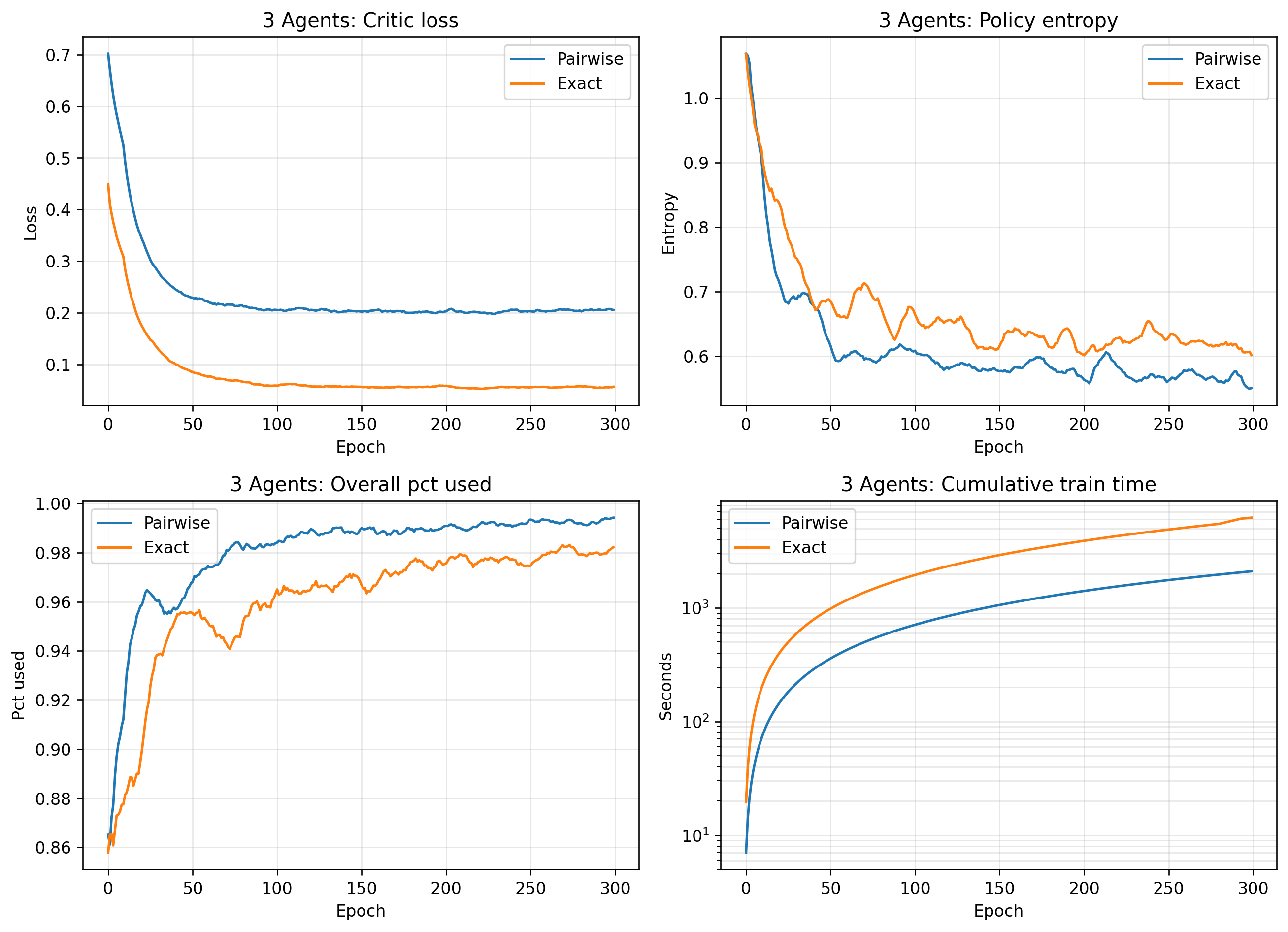}
    \caption{Comparison of exact and pairwise DNQ in the 3-agent bidding game. Exact training achieves lower critic loss, but pairwise training obtains similar or slightly better overall budget usage, lower policy entropy, and substantially lower cumulative training time.}
    \label{fig:three_agent_main}
\end{figure}

Figure~\ref{fig:three_agent_main} compares the exact and pairwise variants of DNQ in the 3-agent setting. The main pattern is clear: the exact shared critic achieves better fit to the full game, but the pairwise shared critic matches or slightly exceeds it on the behavioral metrics while requiring substantially less computation.

First, the exact method reaches a consistently lower critic loss than the pairwise method throughout training. This is expected because the exact critic directly models the full multi-agent payoff tensor, whereas the pairwise critic only captures pairwise interactions. Critic loss therefore measures representational fidelity of the critic, not necessarily the practical quality of the learned bidding behavior.

Second, the two methods are much closer on overall budget usage. Both quickly increase the fraction of budget consumed and converge to high values, indicating that both forms of DNQ learn active and effective bidding policies. In fact, the pairwise shared critic reaches a slightly higher terminal overall usage ratio in this setting, suggesting that the reduced approximation is sufficient to drive strong behavior even without matching the exact critic's reward fit.

Third, the mean policy entropy curves are also close. Both methods become more structured over time, but the pairwise variant converges to slightly lower entropy, indicating a somewhat more concentrated bidding strategy. Since this lower entropy is accompanied by similarly high or higher budget usage, it does not suggest policy collapse; rather, it indicates that the pairwise approximation can still produce decisive bidding behavior.

Finally, the runtime advantage is substantial. Over the full 300-epoch training horizon, the exact method requires approximately three times the cumulative training time of the pairwise method. The exact variant must both predict and solve the full multi-player game at each visited state, whereas the pairwise variant solves only a collection of two-player subgames. The shared critic improves efficiency in both methods by amortizing payoff estimation, but the pairwise formulation still has a clear computational advantage.

\subsection{4-Agent Comparison}

\begin{figure}[t]
    \centering
    \includegraphics[width=\linewidth]{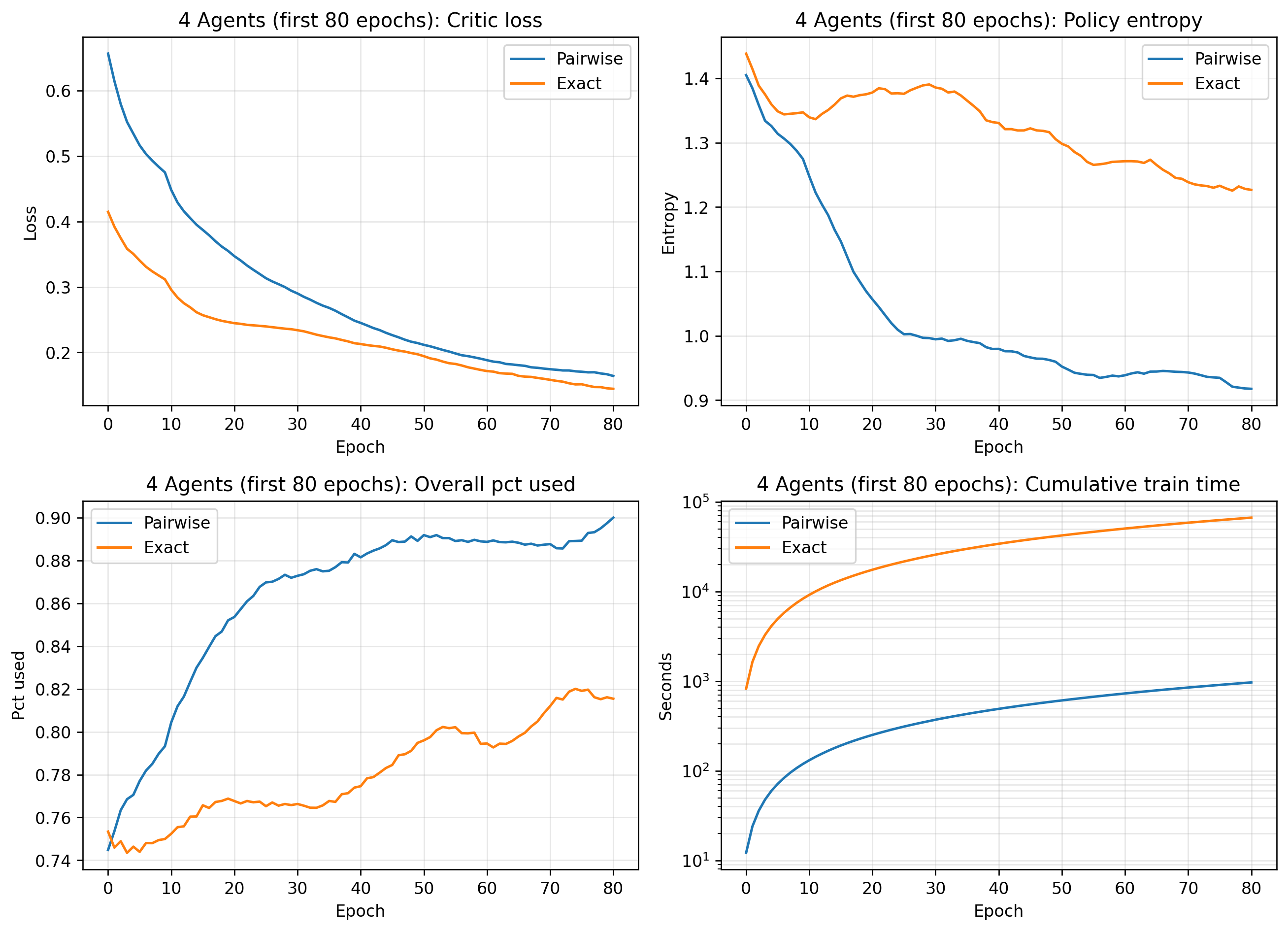}
    \caption{Comparison of exact and pairwise DNQ in the 4-agent bidding game, shown for the first 80 epochs. Pairwise training is dramatically faster and achieves lower entropy and higher overall budget usage earlier, despite having higher critic loss than exact training.}
    \label{fig:four_agent_main}
\end{figure}

The 4-agent setting in Figure~\ref{fig:four_agent_main} makes the scalability gap most visible. Within early training, the pairwise method reduces critic loss from $0.65$ to $0.10$ in 200 epochs using about 40 minutes of training time, whereas the exact method reduces critic loss from $0.41$ to $0.13$ in only 87 epochs but already requires about 20 hours. Therefore, the exact method is roughly $30\times$ more expensive while achieving only a comparable critic-loss scale. The gap grows rapidly because exact DNQ must reason over the full joint payoff structure. By contrast, the pairwise variant continues to operate through pairwise subgames, which are much cheaper to evaluate and solve.

At the same time, the pairwise method performs better on the behavioral metrics in this setting. Its overall budget usage rises faster and reaches a clearly higher level than the exact method within the same training window. The entropy curves support the same conclusion: pairwise entropy decreases more rapidly, indicating that the learned policies become structured earlier. Thus, although the exact critic is more expressive, optimizing it becomes much harder in the 4-agent regime.

Taken together, the 4-agent results provide the clearest evidence for the practical value of the pairwise shared critic: it reduces training cost by a very large margin while producing stronger early behavioral performance.

\subsection{Per-Agent Budget Usage}

\begin{figure}
    \centering
    \includegraphics[width=\linewidth]{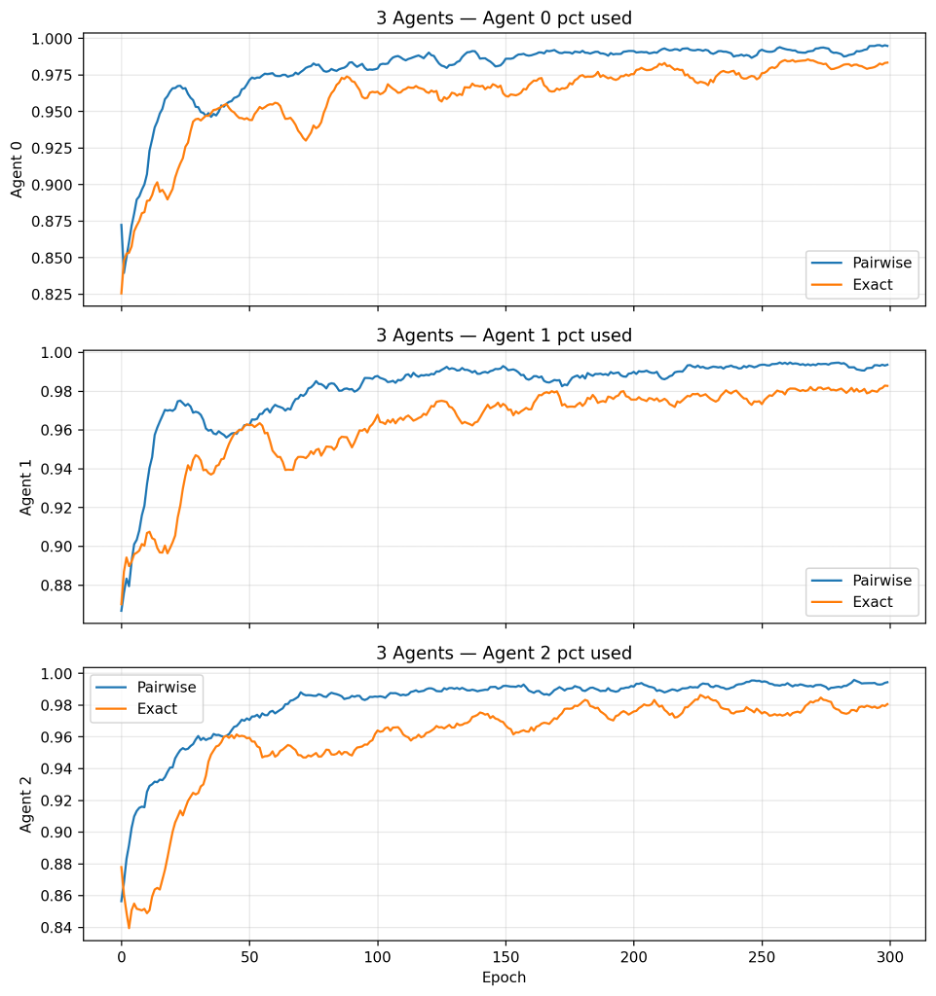}
    \caption{Per-agent budget usage in the 3-agent bidding game. Exact and pairwise DNQ produce similar agent-level consumption patterns, with pairwise typically reaching high usage earlier and maintaining slightly higher utilization across agents.}
    \label{fig:three_agent_pct}
\end{figure}

To examine whether the two methods induce similar behavior at the individual-agent level, Figure~\ref{fig:three_agent_pct} reports per-agent budget usage in the 3-agent setting. The exact and pairwise variants produce broadly similar trajectories for all agents, and both approaches drive all three agents toward high budget usage over time.

The main difference is that the pairwise variant typically reaches high usage earlier and maintains slightly higher usage for each individual agent. This mirrors the aggregate overall budget-usage result and shows that the pairwise approximation does not improve the mean metric by over-optimizing a subset of agents. Instead, it preserves the qualitative structure of the learned bidding behavior across all agents while reducing computation.

This result is also consistent with the use of a shared critic: the learned payoff representation supports coordinated strategic learning across agents rather than inducing highly uneven agent-specific behavior.

\subsection{Discussion}

The experiments show two complementary results. First, in the 2-agent setting, exact DNQ with a shared critic learns stably under direct equilibrium supervision when exact game solving is tractable. Second, in the 3-agent and 4-agent settings, comparing exact and pairwise critics reveals a clear trade-off between critic fidelity and computational efficiency.

The exact method captures the full game structure more accurately, as shown by its lower critic loss, but its computational cost grows quickly with the number of agents. The pairwise method gives up some critic fidelity but remains competitive—and in the 4-agent setting performs better—on key behavioral metrics such as overall and per-agent budget usage and policy entropy.

The main takeaway is therefore:
\begin{quote}
\textit{A shared pairwise critic provides behavior comparable to exact equilibrium supervision in smaller games and a substantially better efficiency--performance trade-off as the number of agents increases.}
\end{quote}

This makes the pairwise variant of DNQ a strong practical alternative to exact equilibrium-guided training for larger multi-agent bidding games.
\section{Conclusion} \label{section:conclusion}

The key contribution of this paper is DNQ, a unified solver-driven training pipeline for equilibrium-informed learning in multi-turn, partially observed multi-agent simultaneous bidding environments. Our results show that while exact equilibrium supervision provides higher critic fidelity, a shared pairwise critic offers a much better efficiency--performance trade-off as the number of agents increases. Future work will study more scalable equilibrium solvers, richer bidding environments, and broader empirical evaluation in larger multi-agent settings.

%
%
%
%
\bibliography{main}
\bibliographystyle{IEEEtran}
%
%

\end{document}